\documentstyle[11pt,epsf,epsfig]{l-aa}

\begin{document}

\thesaurus{03(11.09.4 11.03.2 11.19.3 12.03.3 13.21.1)}

\offprints{D. Kunth (IAP address), kunth@iap.fr}  

\title{HST study of Lyman--alpha emission in star-forming galaxies: 
 the effect of neutral  gas flows 
\thanks{Based on observations with the NASA/ESA {\it Hubble Space Telescope},
obtained at the Space Telescope Science Institute, which is operated by the
Association of Universities for Research in Astronomy, Inc., under NASA
contract No. NAS5-26555.}}

\author{Daniel Kunth\inst{1}, J.M. Mas-Hesse\inst{2}, E. Terlevich\inst{3},
 R. Terlevich\inst{4}, J. Lequeux\inst{5} and S. Michael Fall\inst{6}}

\institute{Institut d'Astrophysique de Paris, 98bis Bld Arago, F-75014 Paris,
 France
\and LAEFF-INTA, POB 50727, E-28080 Madrid, Spain
\and Institute of Astronomy, Madingley Road, Cambridge CB3 0HA, UK
\and  Royal Greenwich Observatory, Madingley  Road, Cambridge CB3 0EZ, UK
\and DEMIRM, Observatoire de Paris, 61 Av. de l'Observatoire, 75014 Paris, 
France 
\and Space Telescope Science Institute, 3700 San Martin Drive, Baltimore,
MD 21218 USA }

\date{Submitted August 1997; Accepted January 1998}

\maketitle

\markboth{Ly$\alpha$\ HST spectroscopy}{}

\begin{abstract}

We present high dispersion HST GHRS UV spectroscopic observations of 8 
H\,{\sc II}
galaxies covering a wide range of metallicities and physical properties. We
have found Ly$\alpha$\ emission in 4 galaxies with blueshifted absorption
features, leading to P~Cygni like profiles in 3 of them.  In all these
objects the O\,{\sc I} and Si\,{\sc II} absorption lines are also blueshifted 
with respect
to the ionized gas, indicating that the neutral gas is outflowing in these
galaxies with velocities up to 200~km\thinspace s$^{-1}$. The rest of the sample shows broad
damped Ly$\alpha$\ absorption profiles centered at the wavelength corresponding to
the redshift of the H\,{\sc II} emitting gas.  We therefore find that the 
velocity
structure of the neutral gas in these galaxies is the driving factor that
determines the detectability of Ly$\alpha$\ in emission.  Relatively small column
densities of neutral gas with even very small dust content would destroy
the Ly$\alpha$\ emission if this gas is static with respect to the ionized region
where Ly$\alpha$\ photons originate.  The situation changes dramatically when most
of the neutral gas is velocity--shifted with respect to the ionized regions
because resonant scattering by neutral hydrogen will be most efficient at
wavelengths shorter than the Ly$\alpha$\ emission, allowing the Ly$\alpha$\ photons to
escape (at least partially).
 This mechanism complements the effect of porosity in
the neutral interstellar medium discussed by other authors, which allows to
explain the escape of Ly$\alpha$\ photons in regions surrounded by static neutral
gas, but with only partial covering factors.  The anisotropy of these gas
flows and their dependence on the intrinsic properties of the violent
star-forming episodes taking place in these objects (age, strength, gas
geometry,...) might explain (in part) the apparent lack of correlation
between other properties (like metallicity) and the frequency of occurence
and strength of Ly$\alpha$\ emission in star-forming galaxies. Attempts to derive
the comoving star--formation rate at high redshifts from Ly$\alpha$\ emission 
searches are highly questionable.

\end{abstract}

\begin{figure*} 
\begin{center}\mbox{\epsfxsize=17cm \epsfbox{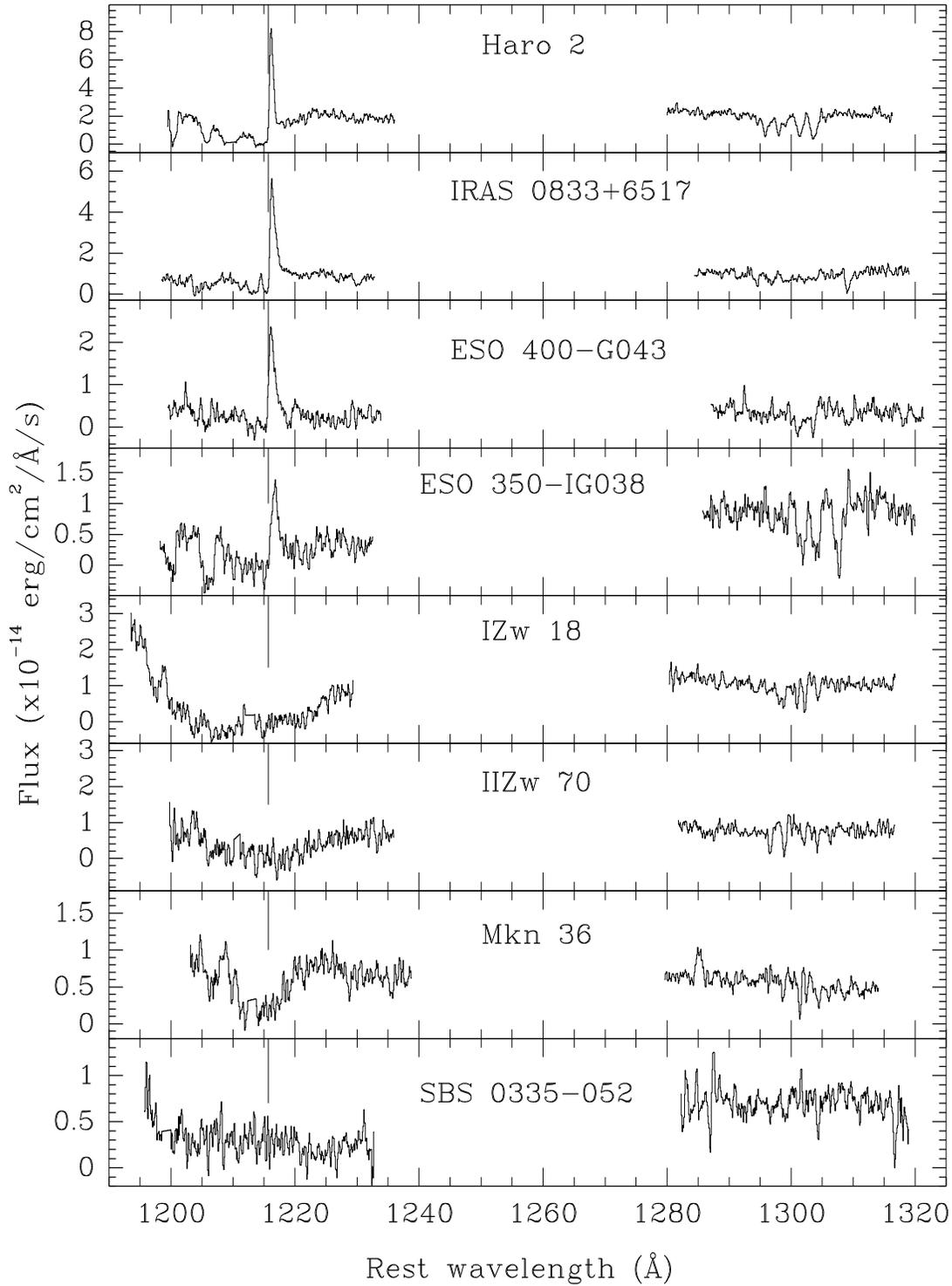}}\end{center}
\caption[]{
GHRS spectra of all the galaxies in the sample. The spectra have been
shifted to rest velocity assuming the redshift derived from optical
emission lines. Vertical bars indicate the wavelength at which the Ly$\alpha$\
emission line should be located. The geocoronal emission profile has been
truncated for the sake of clarity. The spectra have been plotted after
rebinning to 0.1~\AA\ per pixel and smoothed by a 3 pixel box filter.  
}
\label{fig:total}
\end{figure*}

\section{Introduction}

The detection of galaxies at large redshifts that are forming stars for the
first time, the so--called primeval galaxies, remains a very important
astrophysical challenge.  Bearing in mind that galaxy formation may not be
assigned to any preferential cosmological epoch but instead is probably a
continuous process, one might find left--over pristine gas pockets that are
forming young galaxies at the present epoch. For this reason there may be
star--forming galaxies in our local universe that look very much like
distant primeval ones. Hopes have been that Ly$\alpha$\ emission could be a
signature of star formation that would be recognized up to very large
redshifts; hence there have been numerous studies of the
Ly$\alpha$\ emission from distant and local starbursts. Early IUE observations
were performed on more than a dozen nearby starburst galaxies in its SWP
low resolution mode (Meier \& Terlevich 1981; Hartmann et al.
1984; Deharveng et al. 1986; Hartmann et al. 1988 and Terlevich
et al. 1993).  Galaxies with redshifts large enough that their Ly$\alpha$\
emission is separated from the geocoronal line were selected. It was
realized from the very beginning that the Ly$\alpha$/H$\beta$ ratio and the Ly$\alpha$\
equivalent width are much smaller, by at least an order of magnitude, than
expected from the recombination theory. These early works have also shown a
possible anticorrelation between the Ly$\alpha$/H$\beta$ ratio and the H\,{\sc II} 
galaxy
metallicity (actually the O/H abundance, as measured in the ionized gas).

\begin{figure}[t] 
\begin{center}\mbox{\epsfxsize=9cm \epsfbox{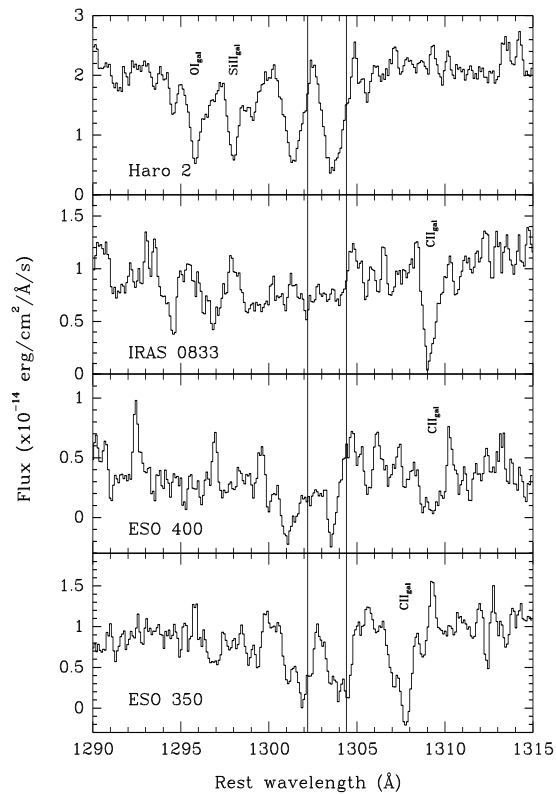}}\end{center}
\caption[]{
Detail on the O\,{\sc I} and Si\,{\sc II} region for the galaxies showing Ly$\alpha$\
emission. The vertical bars indicate the wavelength at which the O\,{\sc I} and
Si\,{\sc II} absorption lines should be located, according to the redshift derived
from optical emission lines. Some Galactic absorption lines have been
marked. Note that the metallic lines appear systematically blueshifted in
these galaxies with respect to the systemic velocity. In some cases there
is no significant absorption at all at zero velocity. 
}
\label{fig:oila}
\end{figure}

\begin{figure}[t] 
\begin{center}\mbox{\epsfxsize=9cm \epsfbox{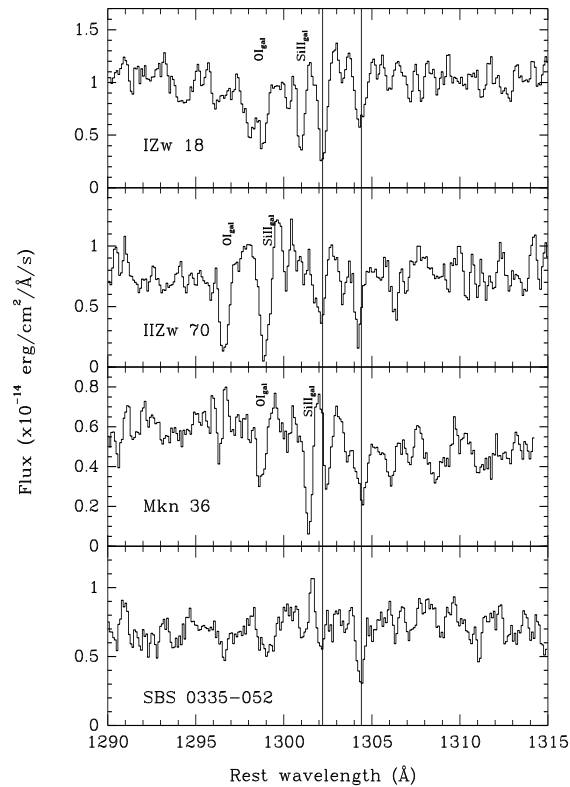}}\end{center}
\caption[]{
O\,{\sc I} and Si\,{\sc II} region for the galaxies showing damped Ly$\alpha$\
absorptions. Details as in Fig.~\ref{fig:oila}. Note that in these
galaxies the metallic lines are essentially at the same redshift than the
ionized gas, indicating the presence of static clouds of neutral gas, as
discussed in the text. 
}
\label{fig:oiab}
\end{figure}

These results, the lack of ``primeval galaxies'' at large redshift in blank
sky searches for redshifted Ly$\alpha$\ emission and the few tentative detections
of Ly$\alpha$\ emission from the damped Ly$\alpha$\ systems have been attributed to the
effects of dust absorption that preferentially destroys Ly$\alpha$\ photons
(Charlot \& Fall 1993, and references therein). The process behind this
is that the transfer of Ly$\alpha$\ radiation is strongly affected by resonant
scattering from neutral interstellar hydrogen atoms. By increasing
enormously their optical path length, Ly$\alpha$\ photons become more vulnerable
to dust absorption, even in small amounts (Neufeld 1991; Charlot \& Fall
1991; Chen \& Neufeld 1994). This process was believed (even in the early
paper of Meier \& Terlevich 1981) to be able to account for the
anticorrelation between the Ly$\alpha$\ emission line visibility and the dust
abundance in these galaxies, as parameterized by the metallicity.
Alternatively such an anticorrelation has been attributed to a 
metallicity--dependent 
extinction law at the wavelength of Ly$\alpha$\ (Calzetti \& Kinney 1992,
Valls-Gabaud 1993). However this conclusion seems very unlikely in view of
the anticorrelation between the Ly$\alpha$\ equivalent width and the gas-phase
abundance of oxygen O/H.  Charlot \& Fall (1993) have emphasized the
advantages of using the Ly$\alpha$\ equivalent widths rather than the
Ly$\alpha$/H$\beta$ ratios, because the former are independent on the extinction
curve of the dust and can be measured with a single observational
device. Their discussion of the anticorrelation between the Ly$\alpha$\ line
equivalent widths and the O/H abundances in a sample of nearby star-forming
galaxies has examined several factors that will affect the observed Ly$\alpha$\
emission from galaxies, among which contributions from supernova remnants
and active galactic nuclei, the orientation of the galaxy and the
absorption by dust. They finally suggest that the structure of the
interstellar medium (porosity and multi-phase structure of the medium) is
most probably the dominant one.

\begin{table*}[t]
\caption{Adopted properties of observed H\,{\sc II} galaxies. Systemic velocities have
been taken from the NASA Extragalactic Database, except for Haro~2 
(Legrand et al. 1997).  } 
  \label{tab:galaxies}
    \begin{tabular}{lcccc}\hline
      Galaxies &  m(V or B) & v(km\thinspace s$^{-1}$) & 12+log(O/H) & E(B-V)  \\ 
      \hline
ESO 350-IG038  & 14.27V & 6156 & ??  & 0.16\\
SBS0335-052    & 16.65V & 4043 & 7.36 & 0.18 \\ 
IRAS 08339+6517& 14.16V & 5730 & ?? & 0.55 \\
IZw 18          & 15.6B  &  740 & 7.17 & $<$0.10 \\
Haro 2         & 13.4V  & 1465 & 8.40 & 0.12 \\
Mkn 36          & 15.5V  &  646 & 7.86 & $<$0.10 \\
IIZw 70         & 14.83V & 1215 & 8.33 & 0.15 \\
ESO 400-G043   & 14.22B & 5900 & 8.0  & 0.20 \\
 \hline
    \end{tabular}
\end{table*}

Our new HST observations indicate that velocity structure in the
interstellar medium plays a key role in the transfer and escape of Ly$\alpha$\
photons. At first place, Ly$\alpha$\ was observed only in absorption in the starburst
dwarf galaxy IZw~18 by Kunth et al. (1994). Since IZw~18 at $Z=1/50
~Z_{\odot}$ is the most metal--poor starburst galaxy known at present, it
was considered previously a good candidate to show Ly$\alpha$\ in emission. To add
to the confusion, a positive Ly$\alpha$\ emission showing a complicated profile,
but a clear P~Cygni component, has been detected in Haro~2, a rather dusty
star-forming galaxy at $Z=1/3 ~Z_{\odot}$ (Lequeux et al. 1995). Giavalisco
et al. (1996) have strengthened the suggestion that the transport of the
Ly$\alpha$\ photons is primarily controlled by the ISM geometry rather than by the
amount of dust, so that the Ly$\alpha$\ emission line would be detected only if
there are holes (regions with low column density of neutral gas) along the
line of sight, a factor which in principle is independent on dust and metal
content of the gas.  As we show hereafter, other factors can be certainly
more important in accounting for variations in the Ly$\alpha$\ emission strength,
at least in some cases. The detection of a P~Cygni profile in the Ly$\alpha$\
emission line of Haro~2 led us to postulate that the line was visible
because the absorbing neutral gas was velocity--shifted with respect to the
ionized gas. This was confirmed by the analysis of the UV O\,{\sc I} and 
Si\,{\sc II}
absorption lines, which were blue--shifted by 200~km\thinspace s$^{-1}$\ with respect to the
optical emission lines, and also of the profile of the H$\alpha$ line
(Legrand et al. 1997).

These new facts and the capability of the HST to analyze in detail for the
first time Ly$\alpha$\ line profiles in nearby galaxies led us to embark on a
longer--term project using the GHRS aiming to study the processes
controlling the detectability of the Ly$\alpha$\ emission line in star-forming
galaxies.  These studies have also been aimed to measure abundances in the
neutral gas of gas--rich dwarf galaxies with spectra dominated by recent
star formation episodes.  Indeed, in objects such as these, the H\,{\sc I} 
clouds
largely extend beyond the optical images suggesting that a substantial
fraction of this gas might still be chemically unevolved or even pristine
(Roy \& Kunth 1995). At a spectral resolution of 20 000 it
became possible to disentangle nebular from stellar absorption lines and to
give crude estimates of the metal abundances in the interstellar medium.
The study of the IZw~18 data by Kunth et al. (1994) and the preliminary
analysis of the rest of the sample (Kunth et al. 1997) have yielded
extremely low values of the O\,{\sc I}/H\,{\sc I} ratios (log N(O\,
{\sc I})/N(H\,{\sc I}) $<$ -7) in some
galaxies of the sample. The complete analysis of the interstellar
abundances will be presented in a forthcoming paper. 

In Sect.~2 we present new HST data on a sample of 8 H\,{\sc II} galaxies. 
Spectra
are described in Sect.~3 and the results are discussed in Sect.~4. 
The conclusions are finally summarized in Sect.~5.

\section{The HST observations}

Eight galaxies have been observed so far, and their properties are listed
in Table~\ref{tab:galaxies}. They were selected by the following
procedure:

\begin{itemize}

\item  The H\,{\sc II} galaxies IZw~18, Mkn~36, IIZw~70 and Haro~2 were first 
chosen
(Cycle~1 and Cycle~4) because they span a wide range of metallicity. The
original aim was to investigate a possible relationship between the
composition of their H\,{\sc II} regions and that of the H\,{\sc I} gas using 
the O\,{\sc I} and Si\,{\sc II} lines. 

\item Results obtained with IZw~18 and Haro~2 prompted us to investigate the 
Ly$\alpha$\ emission profiles per se. Therefore three starburst galaxies were
selected in the IUE-ULDA from the a-priori knowledge that they were Ly$\alpha$\
emitters; they include: IRAS~08339+651, ESO~350-IG038 and
ESO~400-G043. Their redshifts are necessarily larger than those of the
above galaxies because their Ly$\alpha$\ emission on IUE spectra had to be
separated from the geocoronal line.

\item  In addition the SBS~0335-052 spectra, observed by  Thuan et al. (1997) 
with the same setup, were retrieved from the HST archives.

\end{itemize}

Observations were made using the same settings as in Kunth et al. (1994)
and Lequeux et al. (1995) using the Goddard High Resolution Spectrograph
(GHRS) onboard the Hubble Space Telescope (HST). The journal of
observations is given in Table~\ref{tab:journal}. The Large Science
Aperture (LSA) (2 $.\llap"0 \times$ 2 $.\llap"0$) was chosen to ensure a
sufficient flux level. The grating angle was selected according to the
redshift of the objects, so as to cover the Ly$\alpha$\ and the O\,{\sc I} 1302.2~\AA\
regions respectively. The spectral resolution achieved with this setup at
around 1300~\AA\ is close to 0.08~\AA. The Ly$\alpha$\ range was chosen to
investigate both emission and absorption features so that the H\,{\sc I} column
density could be estimated.  The O\,{\sc I} 1302~\AA\ and 
Si\,{\sc II} 1304~\AA\ region was
selected to crudely estimate the chemical composition of the gas and to
measure with reasonable accuracy the mean velocity at which the absorbing
material lies with respect to the star-forming region of a given galaxy.
The spectrum and internal background were moved on the diode array by steps
of one fourth of a diode (GHRS substep pattern 5). In most cases, the
photocathode granularity was averaged out using the GHRS FP--SPLIT = 4
procedure breaking each exposure into four parts between which the grating
is moved by about 5 diodes. We have subsequently extracted those scans to
align and combine them using the standard STSDAS software and form final
spectra with four samples per diode. Wavelength calibrations were achieved
using the platinum-neon lamp\- onboard the spacecraft resulting in an
expected accuracy of the wavelength scale of about 0.08~\AA.  However after
correcting from heliocentric orbital motion of the earth, we noticed on the
IZw 18 spectrum a systematic shift of about 0.24~\AA\ between tabulated
vacuum wavelengths and measured ones. We thus have applied a further shift
so as to match the geocoronal Ly$\alpha$\ line and the observed O\,{\sc I} and 
Si\,{\sc II} lines
originating from Galactic clouds.  We later were informed that the
reduction package had introduced a wrong sign to the heliocentric
correction.  The centroid of the Galactic H~I profile in the direction of
Haro 2 is at -27.5~km\thinspace s$^{-1}$\ LSR, or -23.3 km\thinspace s$^{-1}$\ heliocentric (Hartmann \&
Burton 1995). Therefore we have checked the scale on the Galactic O~I
1302~\AA\
 and Si~\,{\sc II} 1304.4 \AA\ absorption lines, for which we measured
heliocentric radial velocities of -27 and -23 km\thinspace s$^{-1}$\ respectively. Similar
checks with Galactic lines have been performed with the other galaxies in
the sample. Thus the wavelength scale we used should be correct to a few
km\thinspace s$^{-1}$.

\begin{table*}
  \caption{Journal of observations. All spectra obtained through the GHRS
 Large Science Aperture, 
           using the G160M grating, and a 5-fold substep pattern. }
  \vspace*{0.2truecm}
  \begin{tabular}{lcclccr}
\hline
Name  & \multicolumn{2}{c} {Slit position (2000)} & Mode & Date & $\lambda$-range (\AA) &  Exposure  \\
      &             RA      &   Dec       &                 &           &             & time (s) \\
\hline
ESO~350-IG038  & 00 36 52.3 & -33 33 18.2 & FP-SPLIT = NO   &  16/01/96 & 1222 - 1258 &  7018 \\
               &            &             &                 &  16/01/96 & 1312 - 1347 &  4678 \\
SBS~0335-052   & 03 37 44.0 & -05 02 39.0 & FP-SPLIT = DS 4 &  03/01/95 & 1211 - 1247 &  7181 \\ 
               &            &             &                 &  04/01/95 & 1299 - 1335 &  7181 \\
IRAS~08339+6517& 08 38 23.2 &  65 07 15.0 & FP-SPLIT = NO   &  24/02/96 & 1221 - 1257 &  7997 \\
               &            &             &                 &  25/02/96 & 1309 - 1344 &  4787 \\
IZw 18         & 09 34 02.0 &  55 14 27.4 & FP-SPLIT = DS 4 &  23/04/92 & 1195 - 1231 &  9216 \\   
               &            &             &                 &  22/04/92 & 1283 - 1319 & 10137 \\
Haro 2         & 10 32 31.8 &  54 24 03.5 & FP-SPLIT = 4    &  29/04/94 & 1205 - 1241 &  7181 \\ 
               &            &             &                 &  30/04/94 & 1286 - 1321 &  5222 \\  
Mkn 36         & 11 04 58.4 &  29 08 15.2 & FP-SPLIT = 4    &  19/04/95 & 1205 - 1241 &  5984 \\
               &            &             &                 &  20/04/95 & 1281 - 1317 &  5984 \\
IIZw 70        & 14 50 56.5 &  35 34 17.8 & FP-SPLIT = 4    &  08/04/95 & 1204 - 1240 &  3590 \\
               &            &             &                 &  08/04/95 & 1286 - 1321 &  7181 \\
ESO~400-G043   & 20 37 41.9 & -35 29 06.4 & FP-SPLIT = NO   &  16/04/96 & 1221 - 1258 &  7181 \\
               &            &             &                 &  16/04/96 & 1312 - 1347 &  4787 \\
\hline 
\end{tabular}
\label{tab:journal}
\end{table*}

\section {Description of individual spectra}

The individual spectra of all the galaxies in our sample are shown in 
Fig.~\ref{fig:total}.

\subsection{Galaxies with damped Ly$\alpha$\  absorption}

\begin{figure*} 
\begin{center}\mbox{
\epsfig{file=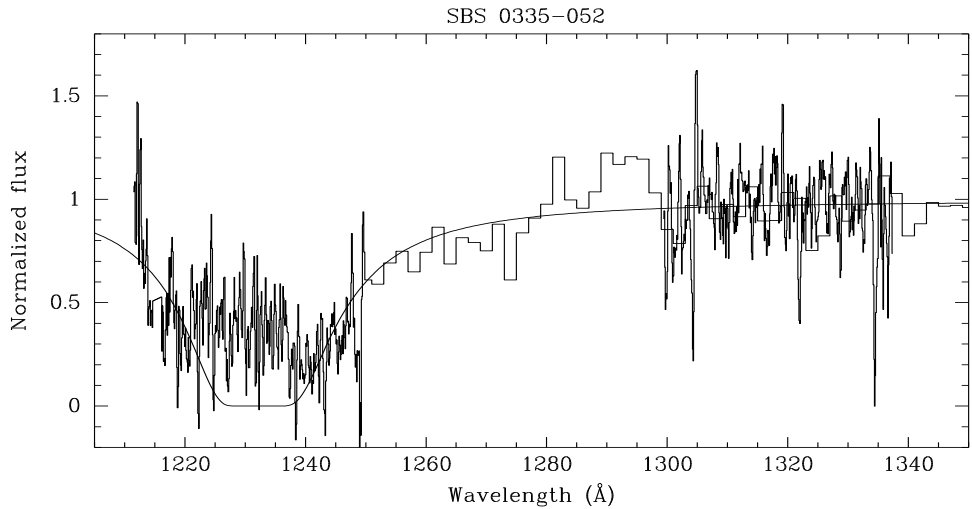,rheight=9cm,height=11cm,angle=-90}}\end{center}
\caption[]{
IUE spectrum of SBS~0035-052 superposed on the GHRS one. Although the GHRS
Ly$\alpha$\ absorption profile is very noisy, its red wing extends clearly into
the IUE range. The Ly$\alpha$\ absorption profile fitted to this red wing is also
shown. 
}
\label{fig:sbs}
\end{figure*}

\begin{itemize}

\item  IZw 18: The HST spectrum shows a damped  Ly$\alpha$\  absorption with no
sign of emission at the redshift of the galaxy (740 km\thinspace s$^{-1}$). This absorption
is a blend of the intrinsic IZw~18 and the Galactic components. A
multi--component fit yields an H\,{\sc I} column density in front of the 
northwest
(NW) emission patch of log~N(H\,{\sc I}) = 21.06 cm$^{-2}$, with a Galactic
component of log~N(H\,{\sc I}) = 20.3 cm$^{-2}$. The multicomponent fit 
requires a
third contribution blueshifted with respect to the Galactic
absorption. This component seems to be an observational artifact due to the
poor signal to noise in the region.  Absorption lines due to O\,{\sc I}
$\lambda$1302~\AA\ and Si\,{\sc II} $\lambda$1304~\AA\ were also detected at the
redshift of the NW H\,{\sc II} region. Unfortunately the O\,{\sc I} line is 
saturated
casting some doubts on any attempts to derive a reliable O/H abundance for
the H\,{\sc I} region (see discussion). O\,{\sc I} and Si\,{\sc II} absorptions
at a velocity of
--160 km\thinspace s$^{-1}$\ due to a Galactic high velocity cloud (No. 117 of Hulsboch \&
Wakker 1988) were detected indicating that the high velocity clouds are not
composed of primordial material.

\medskip 
\item  Mkn 36: A broad  Ly$\alpha$\  is observed in absorption.  
The observed profile can be reproduced by assuming two components: one is
due to neutral gas in Mkn 36 with H\,{\sc I} column density 
of log~N(H\,{\sc I}) = 20.07
cm$^{-2}$ and the second is a Galactic component 
with log~N(H\,{\sc I}) = 19.7. The
O\,{\sc I} region clearly shows O\,{\sc I} and Si\,{\sc II} absorptions that 
are in good agreement
with the systemic velocity of the galaxy. An unidentified emission line or
glitch is seen at 1287.92~\AA . We note that the standard photometric
calibration of the Ly$\alpha$\ region spectrum had to be corrected by an offset of
+2.5$\cdot 10^{-15}$ erg s$^{-1}$ cm$^{-2}$ \AA$^{-1}$, apparently due to poor
background subtraction.

\medskip 
\item  IIZw 70: From the damped Ly$\alpha$\ absorption line we derive
log~N(H\,{\sc I}) = 20.8 cm$^{-2}$, together with a Galactic component with 
log~N(H\,{\sc I})
= 19.3 cm$^{-2}$. Both the Galactic and the intrinsic O\,{\sc I} and 
Si\,{\sc II} lines are
well detected, with the later at the systemic velocity of the galaxy. We
had to offset the Ly$\alpha$\ region spectrum as well by -5.0$\cdot 10^{-15}$ erg
s$^{-1}$ cm$^{-2}$ \AA$^{-1}$ in order to have the core of the absorption
profile at zero level.

\medskip 
\item  SBS~0335-052: The GHRS spectrum of this galaxy has been discussed by 
Thuan et al. (1997). A broad absorption is observed but the intensity of
the central part of the Ly$\alpha$\ profile does not go to zero. Thuan et
al. (1997a) attribute this broad residual emission to resonant scattering
of the Lyman photons that would be re-directed into the line of sight. We
disagree with this interpretation for the following reason:
we noticed that the continuum level is weak (i.e. $<$ 2.0$\cdot 10^{-15}$ erg
s$^{-1}$ cm$^{-2}$ \AA$^{-1}$) hence the GHRS extraction procedure corrects
for an instrumental background level that is a least 10 times higher than
the signal. The signal on this object being the lowest of the sample, the
result is consequently very sensitive to this difficult subtraction and we
suspect the Ly$\alpha$\ region to be spoiled by the extraction procedure. Similar
effects have been discussed above for Mkn~36 (zero level below zero) and
IIZw~70 (zero level above zero), on profiles which otherwise are very well
reproduced by theoretical ones. In this respect it is interesting to note
that the blue wing of the line, which should be dominated by the Galactic
absorption profile, does not reach the zero level either, although the 
H\,{\sc I}
column density along this line of sight is around log~N(H\,{\sc I}) = 
20.68~cm$^{-2}$\
(Dickey \& Lockman 1990). This supports our interpretation about an
instrumental artifact due to low signal to noise ratio. A more recent lower
resolution GHRS spectrum obtained by Thuan \& Izotov (1997) shows indeed
no significant contamination at the core of the absorption line.  

In any case, the red wing of the profile can be extended using the
available IUE spectrum. By combining both spectra, we have fitted the whole
profile up to 1300~\AA, obtaining log~N(H\,{\sc I}) = 21.5 ($\pm 0.2$) cm$^{-2}$, 
as shown
in Fig.~\ref{fig:sbs}. Thuan \& Izotov (1997) obtain log~N(H\,{\sc I}) = 21.8 
cm$^{-2}$\
from their lower resolution spectra.Weak O\,{\sc I} at 1319.66~\AA\ 
(v = 4041 km\thinspace s$^{-1}$)
and Si\,{\sc II} at 1321.94~\AA\ are also detected.

\end{itemize}

\subsection{Galaxies with Ly$\alpha$\  emission}

\begin{itemize}

\item Haro 2: For a full discussion of this GHRS spectrum we refer the
reader to Lequeux et al. (1995). The spectrum around Ly$\alpha$\ is complex.  We
find a deep absorption line at the blue edge, below 1207 \AA\ that is
probably the redshifted NI triplet (1199.6 - 1200.7 \AA) from the
interstellar medium in front of the H\,{\sc II} region of Haro 2. A deep, broad
line at 1211.7 \AA\ is the Si~\,{\sc III} line at 1206.51 \AA\ from Haro 2,
probably interstellar and redshifted by 1260 km s$^{-1}$
(heliocentric). Some Galactic Ly$\alpha$\ interstellar absorption is
detected at zero velocity, with log~N(H\,{\sc I}) = 19.9 cm$^{-2}$. A broad
absorption around 1221 \AA, produced by the gas in front of the star
cluster of Haro 2, is attributed to Ly$\alpha$\ absorption. A strong
asymmetric emission line around 1222.1
\AA\ is Ly~$\alpha$ redshifted by 1580 km s$^{-1}$. The existence of this
line came as a surprise. The spectrum around 1305 \AA\ shows several
absorption lines. Most of the fainter absorption lines are presumably
produced in the stars of Haro~2. The four strong lines are the Galactic
interstellar lines of O\,{\sc I} at 1302.2 \AA\ and Si\,{\sc II} at 1304.4 \AA,
and the
same lines from Haro~2 redshifted by about 1260 km s$^{-1}$. Hence the
heliocentric velocities of the absorption lines are about 200 km s$^{-1}$
lower than the velocity of the H\,{\sc II} region as measured from the H$\alpha$
emission (Legrand et al. 1997). Lequeux et al. (1995) interpreted these
profiles as being produced by a neutral (partially ionized) medium
outflowing from the central star cluster at a projected velocity around 200
km\thinspace s$^{-1}$, as we will discuss later.

\medskip 
\item IRAS 0833+6517: This galaxy has a redshift of
5730~km\thinspace s$^{-1}$. The Ly$\alpha$\ emission measured at 1339.5 \AA\ (flux of 5.6
$\cdot 10^{-14}$ erg s$^{-1}$ cm$^{-2}$ and EW of 34 \AA) is narrow and exhibits
a clear P~Cygni profile. Remarkable enough is a clear secondary emission
situated at 1237.76~\AA\ (at -200 km\thinspace s$^{-1}$\ from the main line component). The
intensity of this secondary peak is 10 times smaller than that of the main
component (with 5$\cdot 10^{-14}$ erg s$^{-1}$ cm$^{-2}$). The absorption
component extents over 1500 km\thinspace s$^{-1}$\ on the blue side of the line
emission. The presence of a secondary emission peak reveals the chaotic
structure of the interstellar medium in this case. Unfortunately no O\,{\sc I}
and/or Si\,{\sc II} absorption is detected that could provide more detailed 
information about the kinematics of the absorbing gas.

\medskip 
\item  ESO-B400-G043: This galaxy is at a redshift of 5900 km\thinspace s$^{-1}$. An
asymmetric Ly$\alpha$\  emission is measured at 1339.5~\AA\ with a flux of
3.1$\cdot 10^{-14}$ erg  s$^{-1}$ cm$^{-2}$ and EW of 20~\AA, showing a P~Cygni
shape. Metallic lines are blueshifted by around -225 km\thinspace s$^{-1}$, with the O\,{\sc I} 
line
at 1326.68 \AA\ (-252 km\thinspace s$^{-1}$) and the Si\,{\sc II} at 1329.18 \AA\ (-194
km\thinspace s$^{-1}$). The absorption profile is best fitted assuming an absorption with
log~N(H\,{\sc I}) of 19.7 cm$^{-2}$, slightly shifted with respect to the metallic 
lines
by -70~km\thinspace s$^{-1}$. There might be an additional secondary emission peak at around
-300~km\thinspace s$^{-1}$, with a flux of 1.1$\cdot 10^{-15}$ erg s$^{-1}$ cm$^{-2}$, but the 
low signal to noise of this spectrum does not allow us to make any firm
conclusion.

\medskip 
\item ESO 350-IG038:  This galaxy has a redshift of 6156 km\thinspace s$^{-1}$.
At this redshift the O\,{\sc I} region falls close to the C\,{\sc II} 
1334~\AA\ Galactic
line which can be used as a reference for the wavelength scale. We find
indeed that the O\,{\sc I} 1302.2 \AA\ and Si\,{\sc II} 1304.4 \AA\ lines are 
at 1328.63
\AA\ and 1330.89 \AA\ respectively, corresponding to a mean velocity of 6097
km s$^{-1}$ or -58 km s$^{-1}$ from the recession velocity of the ionized
regions. In fact both interstellar lines are very broad, indicative of
multicomponents on the line of sight spanning roughly 200 km s$^{-1}$ in
velocity range. A careful inspection of the underlying Ly$\alpha$\ absorption
shows that it extends over more than 1500 km s$^{-1}$ to the blue side of
the emission. The Ly$\alpha$\ emission peaks at 1241.79 \AA\ (its flux is 
1.8$\cdot 10^{-14}$ erg s$^{-1}$ cm$^{-2}$ and EW is 37 \AA ) but does not 
exhibit a
clear P~Cygni profile.  On the contrary, the blue wing of the line does not
sharply drop at zero velocity and moreover the underlying absorption
extends beyond to the red. This agrees with the finding that the metallic
lines are shifted by -58 km s$^{-1}$ with respect to Ly$\alpha$\ but extend to 100
km s$^{-1}$ on both sides. The Ly$\alpha$\ absorption is best fitted with three
components at -26, -197 and -330 km\thinspace s$^{-1}$\ and log~N(H\,{\sc I}) of 18.81, 19.93 and
20.26 cm$^{-2}$, respectively.  The gas coverage is contributed therefore by
numerous components.

\end{itemize}

\begin{table*}
\caption{Measured values for the metallic absorption lines. 
For each object the first line gives the centroid wavelength of the different
lines. The second line gives the systemic velocity, measured from the
optical emission lines, and the corresponding mean velocity offset of the metallic lines
$\delta$v. All wavelengths are given in Angstr\"oms and all velocities in km\thinspace s$^{-1}$.
}

\label{tab:absorptions}
\vspace*{0.2truecm}
\begin{tabular}{lcclccc}
\hline
Name  &  O\,{\sc I}~1302   & Si\,{\sc II}~1304    &     &Galactic &Galactic &Galactic  \\
 v(H\,{\sc II}) km s$^{-1}$ &   v(O\,{\sc I})    & v(Si\,{\sc II})     &$\delta$v &  O\,{\sc I}     & Si\,{\sc II}    &  C\,{\sc II}   \\
\hline
ESO~350-IG038  & 1328.63    &  1330.89    &     & --  & -- &  1334.46 \\
 6156$\pm$31      & 6096.0     &  6099.0     & -58 &    &         &        \\
&&&&&&\\
SBS~0335-052   & 1319.7     &  nd      &     & -- & 1304.29 &  --    \\ 
 4043$\pm$10      & 4030.0  &    --   & -13 &         &         &          \\
&&&&&&\\
IRAS~08339+6517&  nd      &  nd         &     & -- & -- &  1334.15 \\
 5730$\pm$80  &  --      &  --         & -- &         &         &          \\
&&&&&&\\
IZw 18        & 1305.3 &  1307.45    &     & 1301.88 & 1304.06 &  --    \\   
 740$\pm$5  &  721.4     &   708.3     & -25 &         &         &          \\
&&&&&&\\
Haro 2      & 1307.76    &  1310.02    &     & 1302.15 & 1304.34 &  --    \\ 
 1465$\pm$10& 1288.2     &  1299.4    &-171 &         &         &          \\  
&&&&&&\\
Mkn 36         & 1305.3     &  1307.25 &     & 1301.50 & 1304.20 &  --    \\
  646$\pm$5  &  714.5     &   662.3   &  +40 &         &         &          \\
&&&&&&\\
IIZw 70     & 1307.3     &  1309.5     &     & 1301.88 & 1304.17 &  --    \\
  1215$\pm$23 & 1182.2  &  1184.4     & -32 &         &         &          \\
&&&&&&\\
ESO~400-G043   & 1326.7 &  1329.2     &     & -- & -- &  1334.97 \\
  5900$\pm$8 & 5647.0    &  5706.0     &-225 &         &         &          \\
\hline 
\end{tabular}
\end{table*}

We have fitted all absorption
profiles interactively by using the Xvoigt code (Xvoigt,
Copyright 1994, David Mar). In case of damped Ly$\alpha$\ lines blended with the
Galactic line, special weight has been given to the red wing. On the other
hand, when there is an emission feature on the red, the blue side of the
profile and its terminal velocity have allowed to determine precisely the
required H\,{\sc I} column density. In most cases the Ly$\alpha$\ fitting procedure is
insensitive to the $b$ value due to the strong saturation of the profile,
so that only upper limits  have been estimated. 
The line measurements are listed in Table~\ref{tab:absorptions} (metallic
lines) and Table~\ref{tab:lyman} (Ly$\alpha$\ line).  We have also included the
absorption lines attributed to Galactic clouds. We have checked that the
centroid of the Galactic 21~cm line is in good agreement with the velocity
derived from these lines, supporting  our velocity calibration. In
Table~\ref{tab:lyman} we have included the measured logN(H\,{\sc I}) as well 
as the
flux of the emission component and its peak wavelength, if any. In
Fig.5 we show the Ly$\alpha$\ region at rest wavelength with the
fitted H\,{\sc I} absorption profiles superposed to the observed spectra. 

\begin{figure}[t] 
\begin{center}\mbox{\epsfxsize=10cm \epsfbox{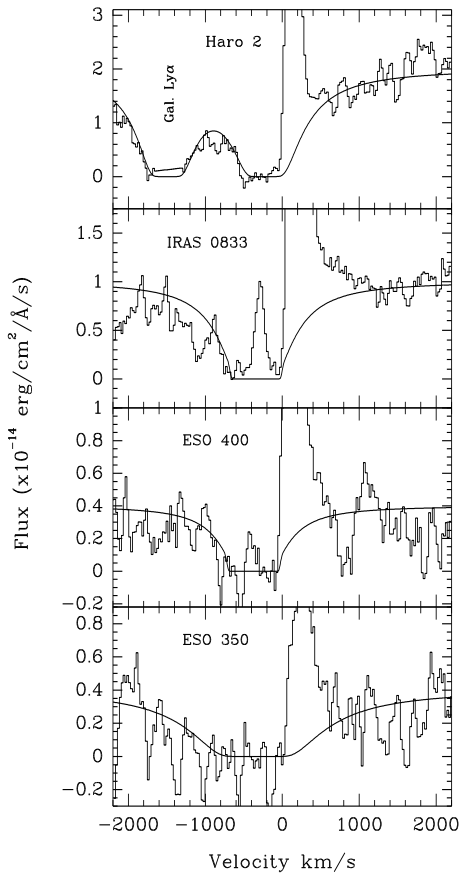}}\end{center}
\caption[]{
Ly$\alpha$\ line profiles plotted in velocity scale. The zero point corresponds
to the systemic velocity as derived from the optical emission lines. The
Galactic Ly$\alpha$\ absorption profile has been also included in 
Haro~2. The geocoronal Ly$\alpha$\ profile has been truncated in this case. The
sharp edge on the blue side of the emission profile is evident in most
cases. Note the secondary Ly$\alpha$\ emission peak in IRAS~0833+6517, and
possibly also in ESO~400-G043.  
}
\label{fig:lalfav}
\end{figure}

\section {Interpretation}

Ly$\alpha$\ photons are produced by recombinations in H\,{\sc II} regions at about 
2/3 of
the ionization rate (the exact yield depends weakly on the density and
temperature of the gas). They are subsequently absorbed and reemitted by H
atoms, both in the H\,{\sc II} regions in which they were produced and in the
surrounding H\,{\sc I} regions, if present. This process -- resonant 
scattering --
changes both the frequency and direction of the Ly$\alpha$\ photons but not their
produced within a galaxy would eventually escape from it, in one direction
or another. This scattering process increases enormously the mean free path
of the trapped photons, so that if some dust is present, the probability of
absorption around the Ly$\alpha$\ wavelength increases also by a significant
factor with respect to the standard UV extinction.  As a consequence,
absorption is potentially important whenever the dust-to-gas ratio exceeds
about one percent of the Galactic value (see, e.g., equation (3) of Charlot
\& Fall (1993)).

If the neutral gas
surrounding the star-forming regions is not static with respect to the
ionized gas, but is outflowing from these regions towards the observer, the
resonant scattering would affect photons at shorter wavelengths than the
Ly$\alpha$\ emission line, i.e., the photons resonantly trapped, and potentially
destroyed by dust, would be mostly stellar continuum photons emitted at
wavelengths below 1216~\AA.  For a galaxy in which the source of ionizing
radiation is a stellar population with a normal initial mass function, the
angle-averaged equivalent width of the Ly$\alpha$\ emission line is about 100~\AA\
in the dust-free case (Charlot \& Fall 1993). This depends only weakly on
the star formation rate in the galaxy provided it is reasonably continuous
(and nonzero over the past few $\times10^7$~yr).  This value can be
somewhat higher if instead the star formation episode is ``instantaneous'',
i.e. lasts less than a few $\times10^6$~yr, as it seems to be the case in
most compact star-forming galaxies. Nevertheless, since the Ly$\alpha$\ photons
would diffuse (in the dust-free case) through the external surface of the
neutral clouds (which are rather large in these compact star-forming
galaxies, extending far beyond the optical regions), its surface brightness
would be very small.  Therefore, even in a dust-free case, we would expect
to detect an absorption line around the Ly$\alpha$\ wavelength if the aperture
sustended by the slit is small compared to the spatial extension of the
neutral cloud. This absorption will be centered at the wavelength
corresponding to the mean velocity of this neutral gas, i.e., it will be
blueshifted with respect to the Ly$\alpha$\ emission line if the neutral gas is
moving towards the observer.  

This scenario allows to explain in a natural
way most of the observational properties in our sample. Among the eight
galaxies observed with the GHRS four show no Ly$\alpha$\ emission at all.
Instead, a strong damped Ly$\alpha$\ absorption at the systemic velocity (as
derived from the optical emission lines) is observed with O\,{\sc I} and 
Si\,{\sc II}
appearing in absorption.  We can infer from Table~\ref{tab:absorptions} and
from the above description that these lines occur without any significant
velocity shift with respect to the H\,{\sc II} regions. This indicates that the
neutral gas in which they mostly originate is static with respect to the
star-forming region. Therefore, since these galaxies have a low dust
content (see Table~\ref{tab:galaxies}; IZw~18 shows weak signs of reddening
and its dust-to-gas ratio is at least 50 times smaller than the Galactic
value - Kunth et al. 1994), this suggests that it remains possible to
observationally weaken Ly$\alpha$\ by simple multiple resonant scattering from the
neutral gas, and even to produce an absorption feature. If this is the
case, the H\,{\sc I} cloud surrounding these galaxies should be leaking Ly$\alpha$\
photons through its external surface. The Ly$\alpha$\ line would then become very
hard to detect because of its low surface brightness. This extended
emission could be detected with deep, large area observations around these
galaxies. Nevertheless, it might be that even the small amounts of dust
present in these galaxies is enough to efficiently destroy a significant
fraction of Ly$\alpha$\ photons, especially if the clouds extension is very
large.  In fact this may be the most inescapable explanation from the
lack of extended Ly$\alpha$\ emission seen at large redshift in blank searches,
since high--z galaxies are expected to have much smaller angular size.

\begin{table*}
  \caption{
Measured values in the Ly$\alpha$\ spectral region. The second line gives the
estimated error bars for logN(H\,{\sc I}), except for ESO~350-IG038 for which the
column density of the three main absorbing components have been
indicated. 
}
 \label{tab:lyman}
  \vspace*{0.2truecm}
  \begin{tabular}{lccccccc}
\hline
Name & $\lambda$(abs) & logN(H\,{\sc I}) &      $b$      & $\lambda$(em.peak)& Flux
& EW &  logN(H\,{\sc I})   \\
      &  \AA          &  cm$^{-2}$ & km\thinspace s$^{-1}$\   & \AA               & erg
s$^{-1}$ cm$^{-2}$ & \AA & Galactic \\
\hline
ESO~350-IG038  & 1240.37    &  20.4   & $<$140 & 1241.9 &  1.8(-14) & 37 &  --  \\
               &            & 18.8--19.9--20.3  &&&&&\\
SBS~0335-052   & 1232.4     &  21.5   & --   & no em. &  --  & --       & --    \\ 
               &            & 21.4--21.7&&&&&\\
IRAS~08339+6517& 1237.7     &  19.9   &  90  & 1239.5 &  5.6(-14) & 34 & --  \\
               &            & 19.7--20.0&    &        &           &&\\
IZw 18         & 1218.6     &  21.1   & --   & no em. &  --         & --  &20.3 \\    
               &            & 21.0--21.5&&&&&\\
Haro 2         & 1220.9     &  19.9   & --   & 1222.1 &  6.0(-14) & 13  &19.8   \\ 
               &            & 19.6--20.5 &&&&&\\
Mkn 36         & 1218.4     &  20.1   &--   & no em. &    --      & --  &19.7 \\ 
               &            & 19.9--20.3&&&&&\\
IIZw 70        & 1220.46    &  20.8   & $<$200 & no em. & -- &  --  & 19.3    \\
               &            & 20.6--21.0&&&&& \\
ESO~400-G043   & 1238.6     &  19.7   &  70  & 1240.0 &  3.1(-14) & 20 & -- \\
               &            & 19.6--19.8&&&&&\\
\hline 
\end{tabular}
\end{table*}

On the other hand, the Ly$\alpha$\ emission in Haro~2 is accompanied by a broad
absorption in the blue wing of that line, with the general appearance of a
typical P~Cygni profile. The amount of neutral gas that produces the blue
absorption trough at Ly$\alpha$\ is rather modest and of the order of
N~(H\,{\sc I}) = 7.7$\cdot 10^{19}$ atom cm$^{-2}$. The crucial point here is that 
the
neutral gas responsible for the absorption in this galaxy is not at the
velocity at which the Ly$\alpha$\ photons were emitted. Moreover, it seems that
all the neutral gas along the line of sight is being pushed by an expanding
envelope around the H\,{\sc II} region, outflowing at velocities close to 200
km\thinspace s$^{-1}$. This interpretation is of course strengthened by the presence of
other detected absorptions of O\,{\sc I}, Si\,{\sc II} and Si\,{SC III} due to 
outflowing gas in
front of the ionizing hot stars of the central H\,{\sc II} region. The 
heliocentric
velocities of all these absorptions are lower by about 200 km s$^{-1}$ than
that of the bulk of the galaxy as measured in the 21-cm line and of the
optical emission lines.  To confirm this hypothesis, Legrand et al. (1997)
have obtained high resolution spectroscopic observations of H$\alpha$ with
the William Herschel telescope at La Palma, finding evidences of an
expanding shell not participating in the rotation of the galaxy.
Comparison of the Ly$\alpha$\ with the H$\alpha$ profiles shows that the Ly$\alpha$\ line
is significantly broader than H$\alpha$, suggesting also scattering of
photons from the back side of the expanding neutral cloud.

The data on the other three H\,{\sc II} galaxies with detected Ly$\alpha$\ emission
confirm that Haro 2 is not an isolated case. All spectra show Ly$\alpha$\ emission
with a broad absorption on their blue side except for ESO 350-IG038 in
which the emission is seen atop of a broad structure requiring several
filaments. When the metallic lines are detected, they are always
blueshifted with respect to the ionized gas, further supporting the
interpretation. In the case of ESO 350-IG038 the velocity structure seems
to be more complicated and several components at different velocities are
identified on the metallic lines. 

The Ly$\alpha$\ absorption profile fitting requires one or several components (in
addition with a Galactic component if the redshift is small). We
find relatively little scatter in the derived column densities (see
Table~\ref{tab:lyman}).  Most clouds have a column density log~N(H\,{\sc I}) of
nearly 19.7 to 21.1. The static clouds tend to have larger column densities
than the moving ones that are also splitted into several components as
expected in a dynamical medium.

The main conclusions that are drawn from this set of data is that complex
velocity structures are determining the Ly$\alpha$\ emission line detectability,
showing the strong energetic impact of the star-forming regions onto their
surrounding ISM. This velocity structure is indeed the driving factor for
the Ly$\alpha$\ line visibility in the objects of our sample.  We want to stress
again that if the absorbing gas is not static with respect to the ionized
region, the Ly$\alpha$\ emission line would be detected, almost independently on
the dust and metal abundance of the gas. It would be affected of course by
the same extinction than the UV continuum but this extinction would not be
enhanced by resonant scattering effects.

If the neutral gas is static with respect to the H\,{\sc II} region, the 
covering
factor by these neutral clouds would probably become the key factor
determining the visibility of the line. Thuan \& Izotov (1997) have indeed
detected strong Ly$\alpha$\ emission in T1214-277, with no evidences of
blueshifted Ly$\alpha$\ absorption. In this case the detection of the line
requires that a significant fraction of the area covered by the slit along
the line of sight is essentially free from neutral gas, suggesting a patchy
or filamentary structure of the neutral clouds. Such a geometry would be
possible only in galaxies not surrounded by enormous H\,{\sc I} clouds, as 
it seems to
be the case in IZw~18 and similar objects. 

The effect of neutral gas flows helps to understand why luminous
high-redshift objects have only been found up to now with linewidths
larger than
1000 km s$^{-1}$.  High--redshift galaxies with very strong (EWs $>$
500~\AA) extended Ly$\alpha$\ emission are characterized by strong velocity shears
and turbulence (v $>$ 1000 km s$^{-1}$); this suggests an AGN activity,
in the sense that other ISM energising mechanism than photoionization by
young stars may be operating. However Steidel et al. (1996) have recently
discovered a substantial population of star--forming galaxies at
3.0$<$z$<$3.5 that were selected not from their emission--line properties
but from the presence of a very blue far-UV continuum and a break below
912~\AA\ in the rest frame. Similarly to our local starbursts they find that
50\% of their objects show no Ly$\alpha$\ emission whereas the rest does, but with
weak EWs no larger than 20~\AA\ at rest.  The Ly$\alpha$\ profiles of this
population look indeed very similar to those of our local starburst
galaxies (Franx et al. 1997 ; Pettini et al. 1997).
We can conclude from the preceeding discussion that the use of Ly$\alpha$\ as a star
formation indicator underestimates the comoving star formation density
at high redshift (e.g. Cowie \& Hu, 1998).

\section{Conclusions}

We have analyzed HST UV spectroscopical data of eight H\,{\sc II} galaxies 
aiming to
characterize the detectability of the Ly$\alpha$\ emission line in this kind of
objects. We obtain the following results:  

\begin{itemize} 

\item Ly$\alpha$\ emission has been observed in four out of the eight H\,{\sc II} 
galaxies. 
In all these four galaxies we have found a clear evidence  of
a wide velocity field by the presence of deep absorption troughs at the
blue side of the Ly$\alpha$\  profiles. Moreover, absorption lines of
metallic elements (O\,{\sc I}, Si\,{\sc II}) are also significantly blueshifted
with respect to the H\,{\sc II} gas velocity.

\item The determining factor for the detectability of the Ly$\alpha$\ emission
line in these galaxies is therefore the velocity structure of the neutral
gas along the line of sight, rather than the abundance of dust particles
alone. If most of the neutral gas is outflowing from the ionized region,
the Ly$\alpha$\ emission line would escape (partially) unaffected, independently
on the metal abundance and dust content of this neutral gas.

This outflowing material apparently powered by massive stars winds and/or
SN may eventually leave the galaxy.  We thus may be witnessing galactic
winds resulting from intense star formation activity. In the case of Haro~2,
Lequeux et al. (1995) suggested that 10$^{7}$ M$\odot$ are expanding at 200
km\thinspace s$^{-1}$.

\item Broad Ly$\alpha$\  absorption is detected in all H\,{\sc II} galaxies. 
The derived N(H\,{\sc I}) column densities lie unexpectedly inside a relatively
small range with 6 of the 8 H\,{\sc II} galaxies having logarithmic column
densities logN(H\,{\sc I}) between 19.9 and 21.1~cm$^{-2}$\
(extreme values are 19.7 and
21.5).  We stress again that the Ly$\alpha$\ photons emitted by the H\,{\sc II}
region are absorbed or redistributed by the H\,{\sc I} gas only if its
velocity is the same as that of  the H\,{\sc II} region. Otherwise,
the photons that are resonantly
trapped were emitted in the stellar continuum close to Ly$\alpha$.

\item The dependence of Ly$\alpha$\ emission detectability on the presence/absence
of neutral static/outflowing gas along the line of sight (and within the
field  of view covered by the slit), helps to explain the apparently
contradictory detection of Ly$\alpha$\ emission in metal and dust--rich galaxies
(like Haro~2), while it may be absent in metal and dust deficient objects,
of which IZw~18 is the prototype. 

\item A partial covering factor of the H\,{\sc II} region by neutral gas, 
with low
H\,{\sc I} column densities, would be required to allow the detection of the 
Ly$\alpha$\ emission line if the neutral gas is static with respect to the ionized
regions.

\item The generally weak or absent Ly$\alpha$\ emission from ``primeval'' and
other galaxies at high redshifts can only be explained by velocity-structure
effects combined with
absorption of the Ly$\alpha$\ photons by dust grains. The relatively small angular
extent of these sources implies that if photons were leaking through the
neutral gas clouds surface after multiple scattering without being
destroyed, the equivalent widths of the lines measured from Earth should be
significantly higher than observed.

\item The present study invalidates attempts to measure the comoving 
star--formation rate density at high redshift on the basis of Ly$\alpha$\ 
emission surveys.

\end{itemize}

Future work should address the several effects discussed in this work
to understand the reasons that govern the presence/absence of the
Ly$\alpha$\ line emission and absorption:  the strength and age of the
burst, the metallicity of the gas (controlling the cooling, hence the wind
evolution), the gravitational potential of the parent galaxy and its
morphology and the H\,{\sc I} and the dust distributions will all play a role. 
The challenge is to determine their relative importance in affecting the Ly$\alpha$\
emission and absorption processes. Clearly the way forward is to realistically 
model the hydrodynamical evolution of the ISM in gas rich dwarf galaxies   
under the influence of starburst of different fractional masses. Particular 
attention should be paid to the time evolution of neutral gas kinematical and 
structural parameters.

\section*{Acknowledgments}

We wish to thank the staff at STScI in Baltimore for his continuous support
during this project.  J.M. Mas-Hesse acknowledges support from Spanish
CICYT through grant ESP95-0389-C02-02. Support for this work was provided
by NASA through grant number GO-05833.01-94A from the Space Telescope
Science Institute. ET and RT acknowledge support from
an EC -- ANTARES -- grant and CNRS respectively during visits to IAP and
LAEFF where part of this work was accomplished.

\label{lastpage}


\begin{thebibliography}{} 
\bibitem{} Calzetti D. , Kinney A.L. 1992, ApJ 399, L39
\bibitem{} Cowie L.L., Hu E.M.,1998,  AJ, in press
\bibitem{} Charlot S. , Fall S.M. 1991, ApJ 378, 471
\bibitem{} Charlot S. , Fall S.M. 1993, ApJ 415, 580
\bibitem{} Chen W.L. , Neufeld N.A. 1994, ApJ 432, 567
\bibitem{} Deharveng J.M., Joubert M. , Kunth D. 1986,
 in the First IAP workshop: ``Star-Forming Dwarf Galaxies and related
objects", edited by. D. Kunth, T.X. Thuan and J. Tran Thanh Van,
 Editions Frontieres, p.431
\bibitem{} Dickey J.M. , Lockman F.J. 1990, ARA\&A 28, 215
\bibitem{} Franx M., Illingworth G.D., Kelson D.D., van Dokkum P.G., Kim-Vy T.
1997, ApJ 486, L75
\bibitem{} Giavalisco M., Koratkar A. , Calzetti D.  1996, ApJ 466, 831
\bibitem{} Hartmann D. \& Burton W.B. 1995, Atlas of Galactic Neutral
Hydrogen (Cambridge University Press, Cambridge)
\bibitem{} Hartmann L.W., Huchra J.P. , Geller M.J. 1984, ApJ 287, 487
\bibitem{} Hartmann L.W., Huchra J.P., Geller M.J., O'Brien P. , Wilson R.
 1988, ApJ 326, 101
\bibitem{} Hulsbosch A.N.M. , Wakker B.P. 1988, A\&AS 75, 191
\bibitem{} Kunth D., Lequeux J., Sargent W.L.W. , Viallefond F. 1994, A\&A 
282, 709
\bibitem{} Kunth D., Lequeux J., Mas-Hesse J.M., Terlevich E. , Terlevich
R. 1997, Rev. Mex. Astr. Astrofis. 6, 61
\bibitem{} Legrand F., Kunth D., Mas--Hesse J.M. , Lequeux J. 1997, A\&A~
326,929
\bibitem{} Lequeux J., Kunth D., Mas--Hesse J.M. ,  Sargent W.L.W. 1995,
A\&A 301, 18
\bibitem{} Meier D.L. , Terlevich R. 1981, ApJ 246, L109
\bibitem{} Neufeld D.A. 1991, ApJ 370, L85
\bibitem{} Pettini M., Steidel C.C,  Adelberger K.L., Kellogg M.,  Dickinson
M.,  Giavalisco M. 1997, to appear in `ORIGINS', ed. J.M. Shull,
C.E. Woodward, and H. Thronson, (ASP Conference Series)
\bibitem{} Roy J.R. , Kunth D. 1995, A\&A 294, 432
\bibitem{} Steidel C.C., Giavalisco M., Pettini M., Dickinson M.,
Adelberger K. 1996 ApJ 462, 17
\bibitem{} Terlevich E., Diaz A.I., Terlevich R., Garcia Vargas M.L.
 1993, MNRAS 260, 3
\bibitem{} Thuan T.X., Izotov Y.I., Lipovetsky V.A. 1997, ApJ 477, 661 
\bibitem{} Thuan T.X. , Izotov Y.I. 1997, ApJ 489, 623
\bibitem{} Valls-Gabaud D. 1993 ApJ 419, 7
\end{thebibliography}
\end{document}